\documentclass[conference]{IEEEtran}
\usepackage{cite}
\usepackage{amsmath,amssymb,amsfonts}
\usepackage{algorithmic}
\usepackage{graphicx}
\usepackage{textcomp}
\usepackage{xcolor}
\usepackage[acronyms]{glossaries}
\usepackage{soul}
\usepackage{siunitx} % for typesetting units
	\DeclareSIUnit\torr{Torr}
\usepackage[caption=false]{subfig}
\usepackage{listings}
\usepackage{array}
\usepackage{matlab-prettifier}
\usepackage{multirow}
\usepackage[hidelinks]{hyperref}
\usepackage{orcidlink}

\PassOptionsToPackage{bookmarks={false}}{hyperref}
\hypersetup{draft}

\def\BibTeX{{\rm B\kern-.05em{\sc i\kern-.025em b}\kern-.08em
    T\kern-.1667em\lower.7ex\hbox{E}\kern-.125emX}}

\newacronym{lidar}{LIDAR}{Light Detection and Ranging}
\newacronym{5g}{5G}{fifth generation}
\newacronym{mmwave}{mmWave}{millimeter wave}
\newacronym{phy}{PHY}{physical layer}
\newacronym{mac}{MAC}{medium access control}
\newacronym{uav}{UAV}{unmanned autonomous vehicle}
\newacronym{em}{EM}{electromagnetic}
\newacronym{iot}{IoT}{Internet of Things}
\newacronym{dl}{DL}{deep learning}
\newacronym{ml}{ML}{machine learning}
\newacronym{drl}{DRL}{deep reinforcement learning}
\newacronym{urc}{URC}{ultra-reliable computing}
\newacronym{urllc}{URLLC}{ultra-reliable low-latency communication}
\newacronym{mimo}{MIMO}{multiple-input multiple-output}
\newacronym{mu}{MU}{multi-user}
\newacronym{rfid}{RFID}{Radio Frequency Identification}
\newacronym{rfp}{RFP}{radio fingerprinting}
\newacronym{sdr}{SDR}{software-defined radio}
\newacronym{mas}{MAS}{Mobile Autonomous System}
\newacronym{rl}{RL}{reinforcement learning}
\newacronym{los}{LoS}{line-of-sight}
\newacronym{dnn}{DNN}{deep neural network}
\newacronym{fpga}{FPGA}{field-programmable gate array}
\newacronym{cv}{CV}{computer vision}
\newacronym{mcs}{MCS}{modulation and coding scheme}
\newacronym{soc}{SoC}{system-on-chip}
\newacronym{mumimo}{MU-MIMO}{\gls{mu}-\gls{mimo}}
\newacronym{dsp}{DSP}{digital signal processing}
\newacronym{snr}{SNR}{signal-to-noise ratio}
\newacronym{csi}{CSI}{channel state information}
\newacronym{svd}{SVD}{singular value decomposition}
\newacronym{cs}{CS}{compressive sensing}
\newacronym{ism}{ISM}{industrial, scientific and medical}
\newacronym{dsa}{DSA}{dynamic spectrum access}
\newacronym{cnn}{CNN}{convolutional neural network}
\newacronym{rfsoc}{RFSoC}{\gls{rf} System on Chip}
\newacronym{ntp}{NTP}{network time protocol}
\newacronym{ptp}{PTP}{precise time protocol}
\newacronym{nu}{NU}{Northeastern University}
\newacronym{tamu}{TAMU}{Texas A\&M University}
\newacronym{ncsu}{NCSU}{North Carolina State University}
\newacronym{uta}{UTA}{University of Texas Austin}
\newacronym{fiu}{FIU}{Florida International University}
\newacronym{uo}{OU}{University of Oklahoma}
\newacronym{gt}{GT}{Georgia Tech}
\newacronym{ucb}{UCB}{University of California Berkeley}
\newacronym{ucsb}{UCSB}{University of California Santa Barbara}
\newacronym{ttu}{TTU}{Texas Tech University}
\newacronym{uh}{UH}{University of Hawaii}
\newacronym{eess}{EESS}{Earth exploration satellite services}
\newacronym{nsf}{NSF}{National Science Foundation}
\newacronym{ntia}{NTIA}{National Telecommunications and Information Administration}
\newacronym{rfc}{RFC}{Request for Comments}
\newacronym{dod}{DoD}{Department of Defense}
\newacronym{frp}{FRP}{foundational research principle}
\newacronym{cct}{CCT}{technological cross-cutting theme}
\newacronym{wiot}{WIoT}{Institute for the Wireless Internet of Things}
\newacronym{ewd}{EWD}{education and workforce development}
\newacronym{pawr}{PAWR}{Platforms for Advanced Wireless Research}
\newacronym{ppo}{PPO}{\gls{pawr} Project Office}
\newacronym{iq}{IQ}{in-phase and quadrature}
\newacronym{if}{IF}{intermediate frequency}
\newacronym{pl}{PL}{physical logic}
\newacronym{lna}{LNA}{low-noise amplifier}
\newacronym{rf}{RF}{radio frequency}
\newacronym{wpan}{WPAN}{wireless personal area network}
\newacronym{wlan}{WLAN}{wireless local area network}
\newacronym{wan}{WAN}{Wide Area Network}
\newacronym{6g}{6G}{sixth generation}
\newacronym{vr}{VR}{virtual reality}
\newacronym{ar}{AR}{augmented reality}
\newacronym{src}{SRC}{Semiconductor Research Corporation}
\newacronym{darpa}{DARPA}{Defense Advanced Research Projects Agency}
\newacronym{adc}{ADC}{analog to digital converter}
\newacronym{dac}{DAC}{digital to analog converter}
\newacronym{di}{DI}{deionized}
\newacronym{gsaps}{GSaps}{Giga-samples-per-second}
\newacronym{awg}{AWG}{arbitrary waveform generator}
\newacronym{dso}{DSO}{digital storage oscilloscope}
\newacronym{nlos}{NLoS}{non-line-of-sight}
\newacronym{thz}{THz}{terahertz}
\newacronym{ghz}{GHz}{gigahertz}
\newacronym{si}{Si}{silicon}
\newacronym{soi}{SoI}{Silicon-on-Insulator}
\newacronym{sige}{SiGe}{Silicon-Germanium}
\newacronym{inp}{InP}{Indium Phosphide}
\newacronym{gan}{GaN}{Gallium Nitride}
\newacronym{gaas}{GaAs}{Gallium Arsenide}
\newacronym{jpl}{JPL}{Jet Propulsion Laboratory}
\newacronym{ic}{IC}{integrated circuit}
\newacronym{ipa}{IPA}{isopropyl alcohol}
\newacronym{hbt}{HBT}{heterojunction bipolar transistor}
\newacronym{hemt}{HEMT}{high-electron mobility transistor}
\newacronym{pa}{PA}{power amplifier}
\newacronym{hdl}{HDL}{hardware description language}
\newacronym{fft}{FFT}{fast Fourier transform}
\newacronym{css}{CSS}{chirp spread spectrum}
\newacronym{dsss}{DSSS}{direct-sequence spread spectrum}
\newacronym{rssi}{RSSI}{received signal strength indicator}
\newacronym{bs}{BS}{base station}
\newacronym{ue}{UE}{user equipment}
\newacronym{nrdz}{NRDZ}{National Radio Dynamic Zone}
\newacronym{ofdm}{OFDM}{Orthogonal Frequency Division Multiplexing}
\newacronym{leo}{LEO}{low Earth orbiting}
\newacronym{trl}{TRL}{technology readiness level}
\newacronym{ldgm}{LDGM}{low-density generator matrix}
\newacronym{ldpc}{LDPC}{low-density parity-check}
\newacronym{lo}{LO}{local oscillator}
\newacronym{isec}{ISEC}{Interdisciplinary Science and Engineering Complex}

\newacronym{osa}{OSA}{OpenAirInterface Software Alliance}
\newacronym{casper}{CASPER}{Collaboration for Astronomy Signal Processing and Electronics Research}
\newacronym{qos}{QoS}{Quality of Service}
\newacronym{oran}{O-RAN}{Open Radio Access Network}
\newacronym{ran}{RAN}{Radio Access Network}

\newacronym{ric}{RIC}{RAN Intelligent Controller}
\newacronym{cbrs}{CBRS}{Citizens Broadband Radio Service}
\newacronym{gaa}{GAA}{General Authorized Access}
\newacronym{pal}{PAL}{Priority Access Licensee}
\newacronym{fcc}{FCC}{Federal Communications Commission}
\newacronym{sas}{SAS}{spectrum access system}
\newacronym{ai}{AI}{artificial intelligence}
\newacronym{ser}{SER}{symbol error rate}
\newacronym{rfic}{RFIC}{radio frequency integrated circuit}
\newacronym{rfi}{RFI}{\gls{rf} Interference}
\newacronym{aml}{AML}{adversarial machine learning}
\newacronym{sdn}{SDN}{software-defined networking}
\newacronym{star}{STAR}{Simultaneous Transmit and Receive}
\newacronym{sinr}{SINR}{signal-to-interference-noise ratio}
\newacronym{vlba}{VLBA}{Very Long Baseline Array}
\newacronym{ngvla}{ngVLA}{Next Generation Very Large Array}
\newacronym{nrao}{NRAO}{National Radio Astronomy Observatory}
\newacronym{fso}{FSO}{Federated Spectrum Observatory}
\newacronym{ngso}{NGSO}{non-geostationary orbit}
\newacronym{vsd}{VSD}{Value-Sensitive Design}
\newacronym{sensr}{SENSR}{Spectrum Efficient National Surveillance Radar}
\newacronym{gbps}{Gbps}{gigabits per second}
\newacronym{tbps}{Tbps}{Terabit-per-second}
\newacronym{nas}{NAS}{Network Attached Storage}
\newacronym{5gb}{5GB}{5G-and-beyond}
\newacronym{osi}{OSI}{Open Systems Interconnection model}
\newacronym{onr}{ONR}{Office of Naval Research}
\newacronym{afosr}{AFOSR}{Air Force Office of Scientific Research}
\newacronym{afrl}{AFRL}{Air Force Research Laboratory}
\newacronym{arl}{ARL}{Army Research Laboratory}
\newacronym{bdss}{BDSS}{broadband directional spectrum sensor}

\newacronym{aoa}{AoA}{Angle of Arrival}
\newacronym{noe}{NOE}{NRDZ Orchestration Engine}

\newacronym{mchem}{MCHEM}{Massive Channel Emulator}
\newacronym{afc}{AFC}{Automated Frequency Coordination}
\newacronym{esc}{ESC}{Environmental Sensing Capability}

\newacronym[firstplural=Devices Under Test (DUTs)]{dut}{DUT}{Device Under Test}

\newacronym{kpi}{KPI}{Key Performance Indicator}
\newacronym{dei}{DEI}{diversity, equity and inclusion}

\newacronym{itu}{ITU}{International Telecommunication Union}
\newacronym[firstplural=Notices of Inquiry (NOIs)]{noi}{NOI}{Notice of Inquiry}
\newacronym{wp}{WP}{Working Party}
\newacronym{drs}{DRS}{Digital Repository Service}
\newacronym{ms}{M.S.}{Master of Science}
\newacronym{apsk}{APSK}{Amplitude and Phase Shift Keying}
\newacronym{hbm}{HBM}{Hierarchical Bandwidth Modulation}
\newacronym{lme}{LME}{Load Modulation Effects}
\newacronym{eirp}{EIRP}{Effective Isotropic Radiated Power}
\newacronym{cte}{CTE}{Continuous-Time Equalizer}

\newacronym{ofdma}{OFDMA}{Orthogonal Frequency Division Multiplexing Access}
\newacronym{otfs}{OTFS}{Orthogonal Time Frequency Space}
\newacronym{papr}{PAPR}{Peak-to-Average Power Ratio}
\newacronym{mf}{MF}{Merit Factor}
\newacronym{mmse}{MMSE}{Minimum Mean Squared Error}
\newacronym{cnt}{CNT}{Carbon Nanotube}
\newacronym{ris}{RIS}{reconfigurable intelligent surface}
\newacronym{xr}{XR}{extended reality}
\newacronym{irs}{IRS}{intelligent reflecting surface}
\newacronym{ap}{AP}{access point}
\newacronym{cw}{CW}{continuous-wave}
\newacronym{thz-tds}{THz-TDS}{terahertz time-domain spectroscopy}
\newacronym{ber}{BER}{bit error rate}
\newacronym{ir}{IR}{infrared}
\newacronym{psg}{PSG}{programmable signal generator}
\newacronym{dsb}{DSB}{double sideband}
\newacronym{vdi}{VDI}{Virginia Diodes, Inc.}
\newacronym{awgn}{AWGN}{additive white Gaussian noise}
    
\begin{document}

\title{Impact of the Antenna on the Sub-Terahertz Indoor Channel Characteristics: An Experimental Approach}

%\title{Sub-Terahertz Path Loss Modeling and Channel Characterization for Ultrabroadband Indoor Links: Examining the Effects of Antenna Gain}

% \title{Sub-Terahertz Channel Measurements for Indoor Deployments: Examining the Antenna Effects}

% \title{Experimental Characterization of Antenna Effects on Ultrabroadband Sub-Terahertz Indoor Channels}

% \title{Experimental Characterization of Antenna Effects on Ultrabroadband Sub-Terahertz Indoor Propagation Channels}

%\title{Experimental characterization of Antenna Gain on Channel Modeling for Sub-Terahertz Ultrabroadband Links


\author{\IEEEauthorblockN{Priyangshu Sen\textsuperscript{*}, Sherif Badran\textsuperscript{\dag}, Vitaly Petrov\textsuperscript{\dag}, Arjun Singh\textsuperscript{*}, and Josep M. Jornet\textsuperscript{\dag}}
\IEEEauthorblockA{\textsuperscript{*}Department of Engineering, SUNY Polytechnic Institute, Utica, NY, USA\\
\textsuperscript{\dag}Department of Electrical and Computer Engineering, Northeastern University, Boston, MA, USA\vspace{-5mm}}

% \author{Priyangshu Sen, Serif Badran, Vitaly Petrov, Arjun Singh, Josep M. Jornet}
% \affiliation{
% \institution{Northeastern University, Boston, MA}}
% \email{j.jornet@northeastern.edu}

% \author{\IEEEauthorblockN{Priyangshu Sen \orcidlink{0000-0002-7618-5908}}
% \IEEEauthorblockA{\textit{Department of Engineering} \\
% \textit{SUNY Polytechnic Institute}\\
% Utica, New York, USA \\
% senp@sunypoly.edu}

% \and

% \IEEEauthorblockN{Arjun Singh \orcidlink{0000-0003-0698-6790}}
% \IEEEauthorblockA{\textit{Department of Engineering} \\
% \textit{SUNY Polytechnic Institute}\\
% Utica, New York, USA \\
% singha8@sunypoly.edu}

% \and

% \IEEEauthorblockN{Sherif Badran \orcidlink{0000-0002-4345-6329}}
% \IEEEauthorblockA{\textit{Department of Electrical and Computer Engineering} \\
% \textit{Northeastern University}\\
% Boston, Massachusetts, USA \\
% badran.s@northeastern.edu}

% \and

% \IEEEauthorblockN{Vitaly Petrov \orcidlink{0000-0002-5235-4420}}
% \IEEEauthorblockA{\textit{Department of Electrical and Computer Engineering} \\
% \textit{Northeastern University}\\
% Boston, Massachusetts, USA \\
% v.petrov@northeastern.edu}

% \and

% \IEEEauthorblockN{Josep M. Jornet \orcidlink{0000-0001-6351-1754}}
% \IEEEauthorblockA{\textit{Department of Electrical and Computer Engineering} \\
% \textit{Northeastern University}\\
% Boston, Massachusetts, USA \\
% j.jornet@northeastern.edu}
}

\maketitle

% % For page numbers to appear, comment before submission
% \thispagestyle{plain} % Comment before submission
% \pagestyle{plain} % Comment before submission

\begin{abstract}
Terahertz-band (100~GHz--10~THz) communication is a promising radio technology envisioned to enable ultra-high data rate, reliable and low-latency wireless connectivity in next-generation wireless systems. However, the low transmission power of THz transmitters, the need for high gain directional antennas, and the complex interaction of THz radiation with common objects along the propagation path make crucial the understanding of the THz channel. In this paper, we conduct an extensive channel measurement campaign in an indoor setting (i.e., a conference room) through a channel sounder with 0.1~ns time resolution and 20~GHz bandwidth at 140~GHz. Particularly, the impact of different antenna directivities (and, thus, beam widths) on the channel characteristics is extensively studied. The experimentally obtained dataset is processed to develop the path loss model and, subsequently, derive key channel metrics such as the path loss exponent, delay spread, and K-factor. The results highlight the multi-faceted impact of the antenna gain on the channel and, by extension, the wireless system and, thus, show that an antenna-agnostic channel model cannot capture the propagation characteristics of the THz channel. 
\end{abstract}


\section{Introduction}
\label{sec:intro}
Sub-terahertz ($100$~GHz--$300$~GHz) communications have been identified among the prospective radio technologies for wireless connectivity in sixth-generation (6G) wireless systems and beyond. The latest predictions account for the rapid development of sub-THz hardware over the recent decade~\cite{thz_hardware_1,thz_hardware_2} and list sub-THz radio among the technologies for 6G sub-THz wireless backhaul~\cite{e_i_backhaul}, 6G sub-THz network sensing~\cite{nokia_sensing}, and 6G sub-THz indoor wireless access~\cite{indoor_thz_li}. Harnessing a new frequency range for an existing scenario typically starts with delivering an understanding of the new wireless channel itself. Here, sub-THz indoor links are not an exception with dozens of measurement and modeling results reported to date~\cite{survey_channels}. 
% focusing on various effects in the sub-THz channels~\cite{survey_channels}.
% ~\cite{priebe2011} ~\cite{chong_l_shape,Christian2011}
For instance, the impact of frequency has been explored in~\cite{xing2021millimeter} among others. The impact of distance has been studied in depth in~\cite{nyu_icc_2022}, and multiple other works. The impact of the environment itself has been explored in multiple studies, including but not limited to~\cite{chong_l_shape}. Admitting the importance and usefulness of prior studies, \emph{in this paper, we primarily focus on the implications of the sub-THz antenna} used for the measurements (and for future sub-THz communication systems). Moreover, we specifically advocate for the following important conclusion with our results: \emph{For 6G and beyond sub-THz radios, there is a clear need to carefully revisit the approach of how a channel model is built}.

% \emph{On the way from 5G-grade mmWave communications to 6G and beyond sub-THz radios, there is a clear need to carefully revisit the approach of how a channel model is built}. 

Specifically, a notable fraction of existing 5G-grade models~\cite{tr_38_901} 
% (where the one in TR 38.901~\cite{tr_38_901} broadly used by the 3GPP is a representative example) 
are built roughly following three steps. First, the measurement campaign is conducted. Then, the measured values get post-processed to compensate for the effect of the used antennas.
% and thus deliver an omnidirectional channel model. 
Finally, there is a method to tailor the delivered model to any desired antenna characteristics. The existing ``antenna-agnostic'' models are very convenient for microwave and even low mmWave systems (e.g., most common 5G New Radio Frequency Range~2 frequencies $24$~GHz -- $30$~GHz). However, extrapolating the same approach to sub-THz bands leads to severe issues during both Stage~2 and Stage~3 above, as \emph{the antenna impact cannot be easily decoupled from the channel impact anymore}. Specifically, as further reported in the paper, some large-scale antennas for sub-THz bands have non-negligible near-field zones of several meters or even tens of meters with complex relationships with the observed channel metrics~\cite{near_field_6g}.
% dependencies between the conditions and the observed channel metrics~\cite{near_field_6g}. 
The occurrence of the waveguide tunneling effect in indoor settings can also have a non-negligible impact~\cite{xu2002spatial}.

% (including ray tracing)
There have been attempts to characterize the impact of antenna directivity with ray-tracing modeling~\cite{priebe2012impact}. However, those techniques have limited applicability in the near field. The ``near-field-to-far-field transition'' when changing the antenna configuration 
% (e.g., from a large antenna to a smaller one) 
leads to a notable deviation in the expected and observed results even in simple line-of-sight (LoS) conditions. Further, sub-THz antennas of different directivity nonlinearly affect the multi-path nature of the received signal, consequently impacting e.g., the delay spread and K-factor in complex ways. Hence, applying the 5G-grade solutions to decouple the antenna-centric effects from channel-centric effects in the measured data, results in notable errors introduced. \emph{To the best of the authors' knowledge, this is one of the first measurement-based studies on antenna-centric effects in sub-THz indoor channels, highlighting and quantifying this important mismatch, as well as suggesting possible ways to partially address the issue.}

The major contributions of this paper are thus as follows:
\begin{itemize}
\item We deliver new measurement results in an indoor configuration with different sub-THz antennas that can be used as reference points for further modeling and performance evaluation efforts in this research area.
\item We further present simple parametric models to characterize the major trends, dependencies, and trade-offs in the observed measurement results. 
% highlighting the deviation from what can be expected when following existing antenna-agnostic models.
\end{itemize}

The rest of the paper is organized as follows. Sec.~\ref{sec:measurement_setup} introduces our measurement setup used for our sub-THz indoor channel sounding. Then, Sec.~\ref{sec:Exp_result} summarizes the major measurement and modeling results, as well as the key observations highlighting the importance and difficulty of incorporating the antenna characteristics into the existing models. Finally, Sec.~\ref{sec:conclusion} concludes the paper and outlines possible further steps in this research direction.

\begin{figure*}
\centering
\includegraphics[width=0.85\textwidth]{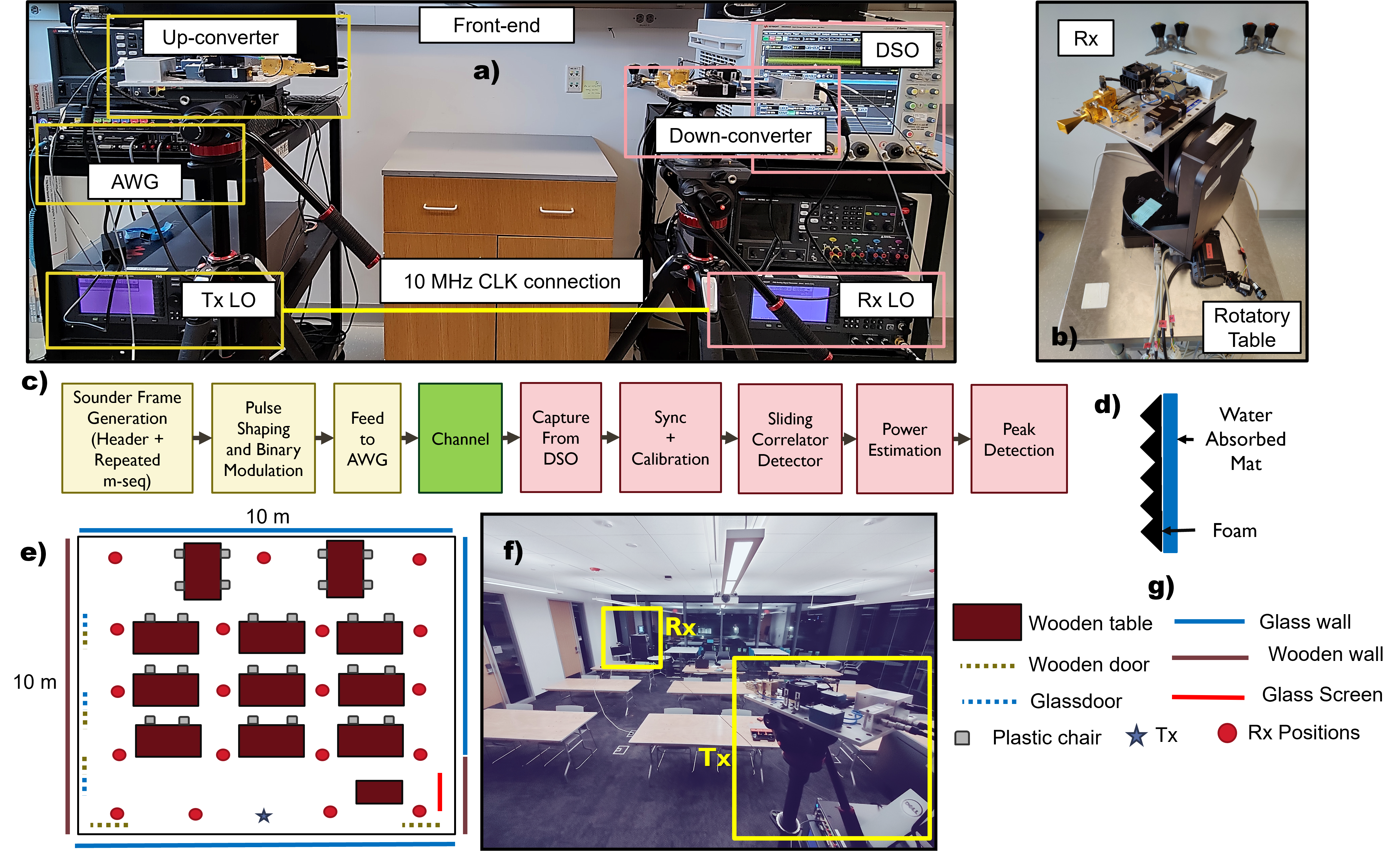}
\caption{The NU Channel Sounder System with a) transceiver hardware, b) rotatory table hardware, c) signal processing backend, and d) THz frequency absorption meterial. The conference room details, i.e., the sounding environment, including e) layout, f) picture, and g) legend.\vspace{-1.2em}}
\label{fig:indoor_ch_sd}
\end{figure*}

\section{Measurement Setup}
\label{sec:measurement_setup}

Herein, we first describe the channel sounder system and then the indoor setup for our measurement campaign.

\subsection{NU Channel Sounder}
\label{subsec:Channel_sounder}

The channel sounder at Northeastern University (NU), or NU Channel Sounder, is a tailored spread spectrum-based sliding correlator type channel sounder~\cite{sen2022terahertz}. It can capture multipath profiles with high resolution and dynamic range.% and consists of a)Terahertz sounder frontends and 2) Signal processing backend.

\subsubsection{Terahertz Sounder Frontends}
\label{subsubsec:Frontend}

The frontend components of the channel sounder are developed through the TeraNova testbed~\cite{sen2021versatile,sen2020teranova}. The transmitter and receiver components, along with the rotatory table, are shown in Fig.~\ref{fig:indoor_ch_sd}~(a) and (b), respectively. At the transmitter, we utilize an arbitrary waveform generator (AWG) to generate baseband (BB)/intermediate frequency (IF) spread spectrum-based channel sounding signals with 20~GHz of bandwidth. Multiplier and mixer chains up-convert the BB/IF to the RF frequency starting from a local oscillator~(LO). 
% at mmWave frequencies. 
The signal is down-converted by a similar RF system and captured through a digital storage oscilloscope (DSO) for offline processing. A cable is utilized to connect the 10~MHz clock of the LOs at the Tx \& Rx for synchronization. An in-home developed absorption material with water-absorbed mat and foam, shown in Fig.~\ref{fig:indoor_ch_sd}~(d), is utilized to mitigate the reflection from the metal structure of the receiver itself.

\subsubsection{Signal Processing Backend}
\label{subsubsec:Backend}

The signal processing backend consists of both transmitter and receiver sections, as shown in Fig.~\ref{fig:indoor_ch_sd}(c), and is implemented in MATLAB. 
We utilize the backend to generate the channel-sounding specific waveform and further process it at the receiver.
% the BB/IF signal samples at the transmitter. We also process the received signal to acquire the channel impulse response or multipath profile. 
% The different signal processing blocks of the backend are developed using MATLAB. 
The transmitted signal frame is made of an 8191-m-sequence header and a 4095-chips m-sequence that is repeated 16 times, which is utilized for actual channel measurement. The frame is modulated with binary phase-shift keying (BPSK) and pulse-shaped with a high roll-off factor root-raised cosine filter. The signal is generated through the AWG and up-converted for transmission to the 130-150 GHz band.
At the receiver, the captured sampled signal at the DSO is processed, by first utilizing the header to detect the start of a frame. Then, the captured signal is calibrated to eliminate the frequency selective response of the hardware as shown in~\cite{sen2022terahertz}. 
% For this purpose, the frequency response of the aforementioned devices are measured as
% \begin{equation}\label{eq:calibration}
% H(k)=\biggl(\frac{P_r (k)-P_n}{P_s (k)}\biggr)^\frac{1}{2},
% \end{equation}
% where $P_r(k)$ is the received signal power with noise at the $k_{th}$ frequency, $P_s(k)$ refers to the transmitted signal power, and $P_n$ is noise power for the entire observation bandwidth. To mitigate the frequency selectivity of the hardware from received signals, the inverse of normalized $H$ is used. 
The signal-sounding component of the waveform (with repeated 4095 m-sequence) is correlated with a locally generated 4095 m-sequence to extract the channel impulse response. An average of over 16 such repetitions are taken to reduce the noise floor and eliminate false detection.
The received power is determined by normalizing with pulse energy and m-seq length (i.e., 4095). 
% We average the impulse response over 16 such repetitions in order to reduce the noise floor and eliminate false detection.
% The received power is determined by normalizing the processed channel impulse response with pulse energy and m-seq length (i.e., 4095). 
Finally, by setting the threshold based on the channel sounder's dynamic range, the peaks, and corresponding delays are measured by the global convergence method. With 20~GHz of RF bandwidth, multipath components with 0.1~ns time resolution can be measured.

\subsection{Sounding Environment and Methodology}
\label{subsec:sounding_env}

The measurement campaign is performed in an indoor conference room of dimension 10m$\times$10m at Northeastern University. The conference room structure, geometrical shape, and placement of the transceiver are shown in Fig.~\ref{fig:indoor_ch_sd} (e),(f) and (g). We have utilized 13~dBm of transmit power with a 15~dBi ($30^{0}$ beamwidth) antenna to transmit the sounding signal at a height of 2~m close to the ceiling. 
The receiver is kept at the height of 1~m to define the height of the tabletop on the rotatory table in the different places within the room, as shown in Figure~\ref{fig:indoor_ch_sd} (e).
% (portrayed as the red circles in Fig.~\ref{fig:indoor_ch_sd} (e)), which are 2~m apart from each other. 
% For this purpose, we divided the room into grids of $2 \times 2 m^{2}$. 
The rotatory table is rotated in steps of the beamwidth of the receiving antenna in the azimuth direction to acquire the angular information. We start by ensuring that our transmitter and receiver antennas are perfectly aligned, maintaining LoS. At the receiver, antennas of various gain/beam widths are utilized to collect the received signal. The antennas we have utilized have a gain of 15, 21, 25, and 38 dBi, and beamwidths of 30, 11, 10, and $2^{\circ}$, respectively. All the antennae at the transceiver operate at the D-band frequency range (110-170~GHz).

\section{Experimental Results}
\label{sec:Exp_result}

First, we analyze the path loss model and exponent with various antenna gains.
% for numerical analysis, which is crucial for link budget analysis, and evaluate the link quality. 
Finally, we moved with the detected multipath profiles and statistically determined channel metrics. 
% such as K-factor, delay spread, and angular spread.

\subsection{Path Loss Model}

The path loss models are crucial to estimate power loss over the wireless interface due to multipath interference. 
% i.e., the constructive or destructive superposition of multipath components. 
% To this end, it is crucial to evaluate the link budget within the system design, from which the antenna gains can be removed and the subsequent path loss can be evaluated. 
The presence of a line-of-sight (LoS) component normally results in representing the path loss through the log-distance model. Here, the path loss $\mathrm{PL}$ at a distance $d$ is evaluated through the Friis free space path loss $\mathrm{PL}_{0}$ at $d_0 = 1~m$, with the excess loss due to the distance, $d$, having a path loss exponent (PLE), $n$. A zero-mean Gaussian distributed random variable, $\chi$, with a standard deviation of $\sigma$~dB, is introduced to represent the shadow effect. The cumulative path loss $\mathrm{PL}$ is given by 
\begin{equation}\label{eq:pathloss_indoor}
\mathrm{PL}=\mathrm{PL}_{0} + 10 n \log_{10} \left( \frac{d}{d_{0}} \right) + \chi.
\end{equation}

\begin{figure}
\centering
\includegraphics[width=0.9\columnwidth]{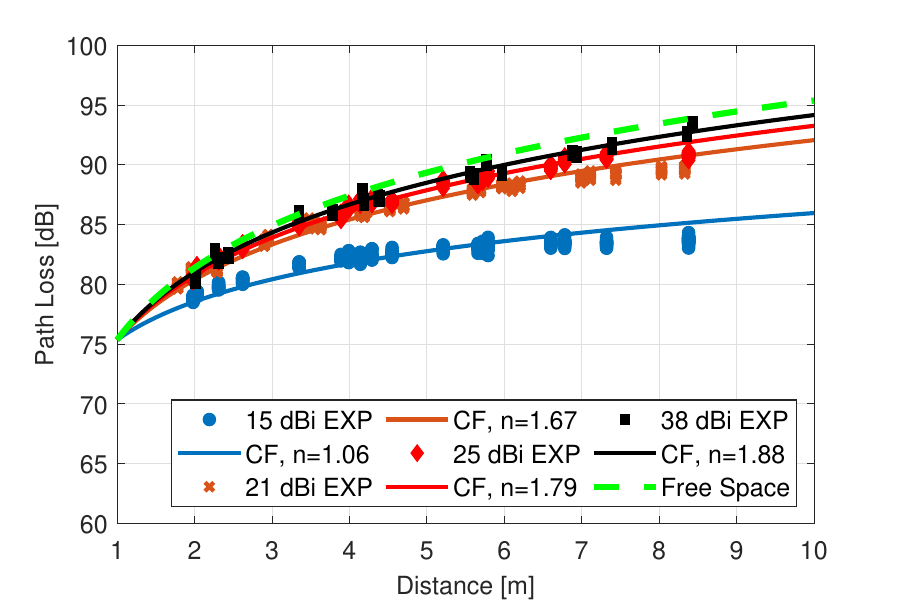}
\caption{Path Loss with varying antenna gain (beam width) with experimental (EXP) and curve fitted (CF) value.\vspace{-1.2em}}
\label{fig:path_loss_in}
\end{figure}

\begin{figure*}
\centering
\subfloat[Path Loss Exponent]{\includegraphics[width =0.33\textwidth]{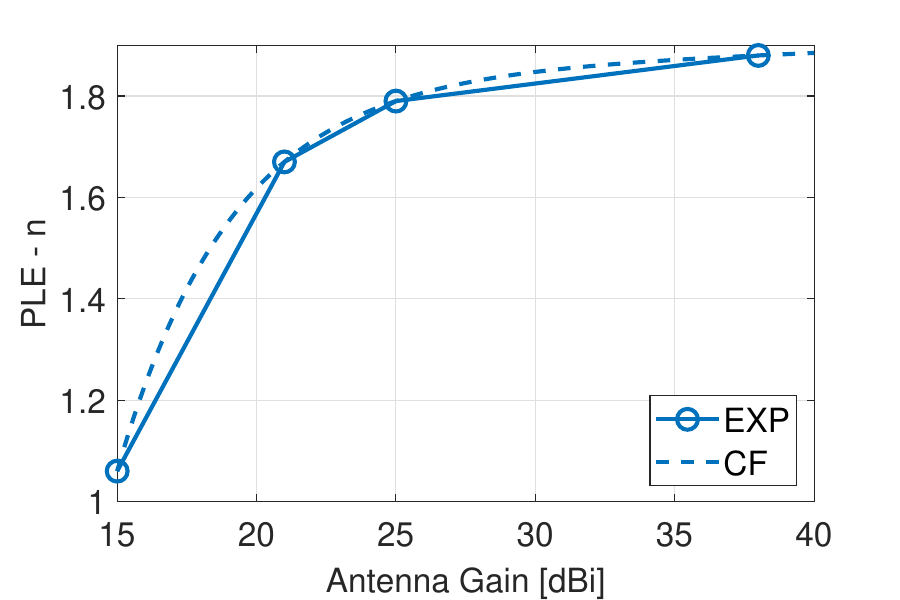} \label{fig:CF path loss exponent}}
\subfloat[K-factor]{\includegraphics[width =0.33\textwidth]{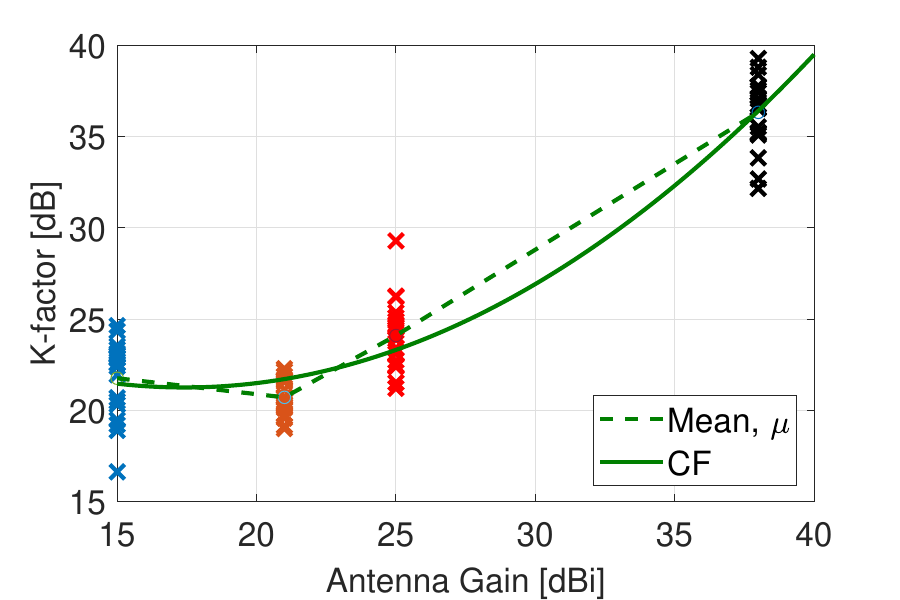} \label{fig:CF K-factor}}
\subfloat[RMS Delay Spread]{\includegraphics[width =0.33\textwidth]{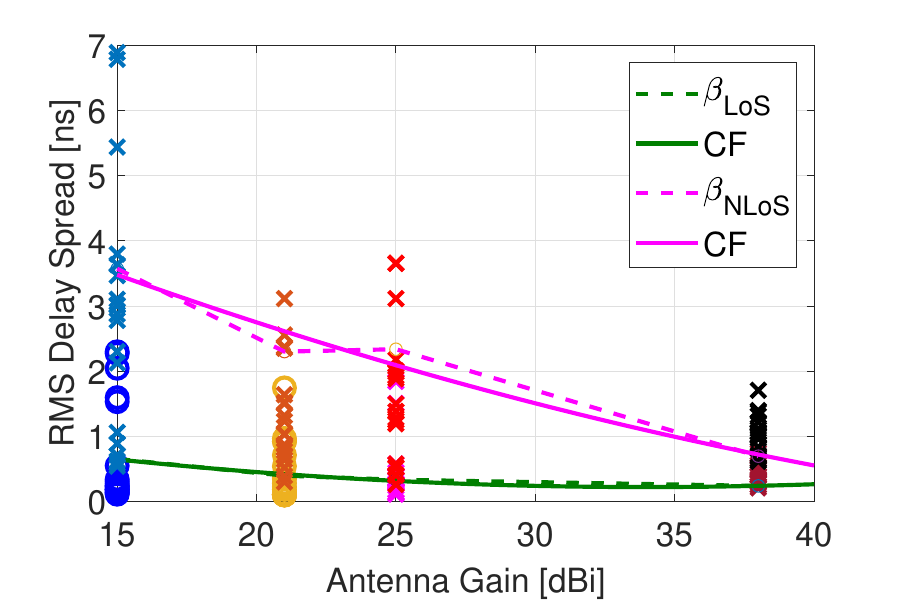} \label{fig:CF RMS DS}}
\caption{Experimental and curve-fitted channel parameters with varying antenna gains for the conference room.\vspace{-1.2em}}
\label{fig:CF_parameter}
\end{figure*}

We estimate the value of $n$ by mapping the experimental path loss, $L_\mathrm{PL}$, in dB scale using the maximum likelihood estimation (MLE) method with the model given in~\eqref{eq:pathloss_indoor} for the link budget analysis. 
%
% $L_\mathrm{PL}$ is found through the link budget analysis:
% \begin{equation}\label{eq:link_budget}
% L_\mathrm{PL}=P_\mathrm{tx}+G_\mathrm{tx}+G_\mathrm{rx}+G_\mathrm{LNA}-L_\mathrm{mixer}-L_\mathrm{misc}-P_\mathrm{rx},
% \end{equation}
% where $P_\mathrm{tx}$ is the transmitted signal power, $G_\mathrm{tx}$ and $G_\mathrm{rx}$ are the transmit and receive antenna gains, respectively. $G_\mathrm{LNA}$ is the low-noise amplifier (LNA) gain at the receiver. $L_\mathrm{mixer}$ is the conversion loss at the receiver and $L_\mathrm{misc}$ accounts for miscellaneous losses in cables and connectors. 
%
Figure~\ref{fig:path_loss_in} presents both the experimental as well as estimated/curve fitted (CF) values of path loss as a function of distance for different antenna gain at the receiver. 
% The $L_\mathrm{PL}$ is estimated by using link budget analysis and represented by
% \begin{equation}\label{eq:link_budget}
% L_\mathrm{PL}=P_\mathrm{tx}+G_\mathrm{tx}+G_\mathrm{rx}+G_\mathrm{LNA}-L_\mathrm{mixer}-L_\mathrm{misc}-P_\mathrm{rx},
% \end{equation}
% where $P_\mathrm{tx}$ is the transmitted signal power, $G_\mathrm{tx}$ and $G_\mathrm{rx}$ are the transceiver antenna gains. $G_\mathrm{LNA}$ is low noise amplifier gain, and $L_\mathrm{mixer}$ is the conversion loss at receiver. $L_\mathrm{misc}$ accounts for miscellaneous losses in cables and connectors. 
% The estimated/curve fitted (CF) value and experimental value of path loss are plotted with distance in Fig.~\ref{fig:path_loss_in} for different antenna gains. 
% \begin{table}
% \caption{Path Loss exponent, n, and $\chi$  statistics for different antenna gain with transmit antenna gain of 15~dBi}
% \centering
% \begin{tabular*}{\columnwidth}{@{\extracolsep{\fill}}cccc}
% \toprule
% Antenna & $\mu$ & $\sigma$ & $n$ \\
% \midrule
% 15 dBi ($30^{\circ}$) & 0 dB & 0.72 dB & 1.06 \\
% 21 dBi ($11^{\circ}$) & 0 dB & 0.54 dB & 1.67 \\
% 25 dBi ($10^{\circ}$) & 0 dB & 0.47 dB & 1.79 \\
% 38 dBi ($2^{\circ}$) & 0 dB & 0.45 dB & 1.88 \\
% \bottomrule
% \end{tabular*}
% \label{tab:PL_ant}
% \end{table}
Figure~\ref{fig:CF path loss exponent} shows the variation of the path loss exponent with varying antenna gains. The curve fit is given by
\begin{equation}
    n(G) = 1.811e^{0.001018G} -30.15e^{-0.2437G},
\end{equation}
where $n$ is the PLE and $G$ is the antenna gain. The value of $n<2$ in most cases is caused by the waveguide/tunnel multi-path effect of LoS indoor channel~\cite{xu2002spatial}. Further, $n$ increases with an increase in antenna gain (i.e., a decrease in antenna beamwidth). This increase in path loss with antenna gain can not be explained alone by the lowering of the waveguide/tunneling effect as shown in~\cite{priebe2012impact} through ray tracing simulation. 
% Specifically, increasing the gain requires increasing the antenna size proportionately, which increases the distance at which the far-field assumption becomes valid. This additional increase in the path loss can be understood as 
This can be more clearly understood as the fact that the effective gain of larger gain, and consequently larger sized aperture, is reduced due to the presence of near-field effects. Specifically, we can estimate the radiated and received beams from these antennas by approximating them as Gaussian beams~\cite{alda2003laser}. We quantify the evaluations for the extreme cases of 15 and 38~dBi antennas, with the other antennas falling in this range. The 38~dBi antenna is a circular horn-lens antenna (efficiency of 0.45), giving us a Gaussian beam with an aperture $w_{0, 38} = 59$~mm~\cite{balanis2016antenna}. In contrast, the 15~dBi antenna is an open waveguide, with the effective gain given as~\cite{balanis2016antenna}: 
\begin{equation}
    G = 10 \log_{10}\left( 0.81 \frac{4\pi}{\lambda^{2}} ab \right),
\end{equation}
where $\lambda$ is the wavelength, and $ab$ is the product of the width and length of the waveguide. The correction factor of 0.81 comes as $\approx 8/\pi^2$, due to the fact that the electric field across the waveguide is not uniform. We equate this waveguide to a circular aperture with the same efficiency of 0.45, giving the value as $w_{0, 15} = 4$~mm. 

\begin{figure*}
\centering
\includegraphics[width=\textwidth]{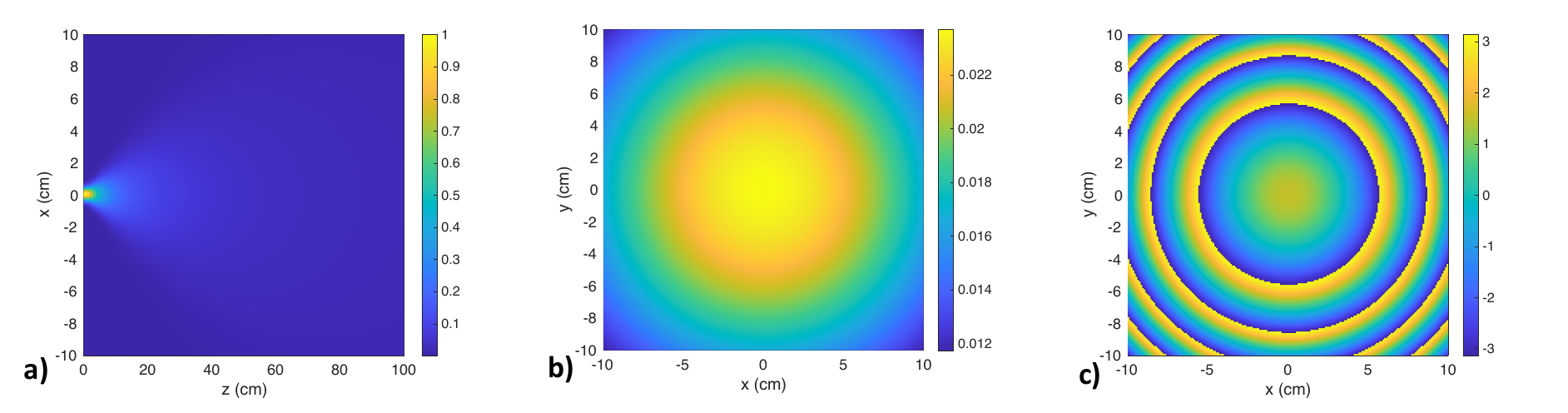}
\caption{Beam profile from a 15~dBi antenna: (a) propagation; (b) cross-sectional amplitude; and (c) cross-sectional phase. The cross-sectional cut is at 1~m.\vspace{-1.2em}}
\label{fig:gain_phase_indoor}
\end{figure*}

A Gaussian beam generated from an aperture $w_0$ with an initial electric field $E_0$ has the field $E(z)$ after propagation in the z-axis given as~\cite{alda2003laser}: 
\begin{align}
\label{eq:electric_field}
&E(z) = E_{0}\frac{w_0}{w(z)}e^{\left( \frac{-r^2}{w(z)^2} e^{\left( -j(kz + k\frac{r^2}{2R(z)} + \phi(z))\right)}\right)},\\ 
&\mathrm{where}\quad{} w(z) = w_{0}\sqrt {1 + \left(\frac{z}{z_{R}}\right)^{2}}, \quad{} R(z) = \frac{z^2 + z_{R}^{2}}{z}\nonumber,
\end{align}
where $w(z)$ is the beam waist after propagating a distance $z$, and $R(z)$ is the radius of curvature, with $\phi(z) = \arctan{z/z_{R}}$ describing the Gouy phase. $z_{R} = \pi w_{0}^{2}/\lambda$ is the Rayleigh range. The received power $P_\mathrm{rx}$ from a receiver with an aperture of size $w_\mathrm{rx}$ at a link distance $z$ is found as:
\begin{equation}
   P_\mathrm{rx} = \int_{-w_\mathrm{rx}}^{w_\mathrm{rx}} E(z)^{2} \,dr,
\end{equation}
where $E(z)$ is the electric field from~\eqref{eq:electric_field}. The phase component $-k\frac{r^2}{2R(z)}$, where $k$ is the wavenumber, varies across the cross-sectional aperture $w_\mathrm{rx}$, and thus the electric field does not add up in phase. We present the beam profile emanating from the 15~dBi transmitter in Fig.~\ref{fig:gain_phase_indoor}. The cross-sectional cut is at $z=1~m$, similar to how $\mathrm{PL}_0$ is set in~\eqref{eq:pathloss_indoor}. It can be seen that the beam intercepted by a smaller receiver will have a uniform phase profile, whereas the larger antenna will capture a beam that has multiple variations in the phase, indicating that there will be significant destructive interference, reducing the operational gain, or equivalently, increasing the path loss. 

Nonetheless, the deviations of each of the experimentally evaluated curves from each other as well as the conventional free-space path loss model highlight that channel characterization in the THz regime cannot be performed independently of the antenna(s) utilized in the system.

\subsection{Channel Metrics}
\label{subsec:metrics}
We characterize and present here the ultra-broadband indoor wireless communication link at 140~GHz in terms of the Rician K-factor, the root mean square (RMS) delay spread (DS), and the angular spread (AS), which are among the crucial metrics for the over the air communication system design. In addition, we have demonstrated the relationship between the metrics and antenna gain/beamwidth.

\subsubsection{K-factor}
\label{subsubsec:k_factor}

\begin{figure}
\centering
\includegraphics[width=0.9\columnwidth]{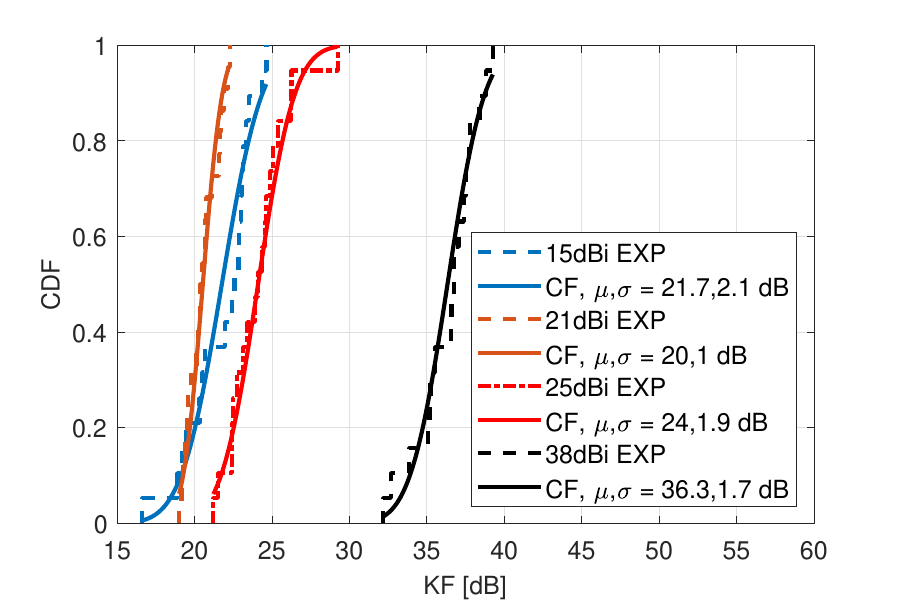}
\caption{CDF of K-factor with different antenna gain for EXP result (dotted line) and its CF (solid line) with $\mu$, and $\sigma$ of log-normal distribution.\vspace{-1.2em}}
\label{fig:Kf_indoor}
\end{figure}

The K-factor is a significant metric for LoS link, which provides insight into communication link quality based on the power associated with the LoS and the scatter components. It is defined by the ratio of the power in the LoS path ($P_\mathrm{LoS}$) and other NLoS paths ($2~S^{2}$), and it thus given~by 
\begin{equation}\label{eq:k_factor_in}
K=\frac{P_\mathrm{LoS}}{2S^{2}}.
\end{equation}
In the dB scale, it is represented by $10\log_{10}(K)$. This metric provides insight into the type of fading a channel could experience. For example, with an increase in the K-factor, the chance of experiencing deep fade reduces and increases the link reliability by reducing the bit error performance. Therefore, the estimation of the K-factor is of practical importance in various wireless scenarios, including channel characterization, adaptive modulation, and localization applications. 
% Further, it helps to decide on suitable modulation and coding schemes to increase the link reliability.

% \begin{figure}
% \centering
% \subfloat[Antenna gain]{\includegraphics[width =0.5\columnwidth]{Figures/KF_ant_BW.pdf}}
% \subfloat[Occupancy]{\includegraphics[width =0.5\columnwidth]{Figures/KF_occupancy.pdf}}\\
% \subfloat[Room size]{\includegraphics[width =0.5\columnwidth]{Figures/KF_room_area.pdf}}
% \caption{Statistics of K-factor at different scenarios.}
% 	\label{fig:Kf_indoor}
% \end{figure}

% \begin{table}
% \caption{K-factor statistics for different antenna gain with transmit antenna gain of 15~dBi}
% \centering
% \begin{tabular*}{\columnwidth}{@{\extracolsep{\fill}}cccc}
% \toprule
% Antenna & $\mu$ & $\sigma$ & CDF Range \\
% \midrule
% 15 dBi ($30^{\circ}$) & 21.75 dB & 2.05 dB & 16-24 dB \\
% 21 dBi ($11^{\circ}$) & 20.50 dB & 0.99 dB & 19-22 dB \\
% 25 dBi ($10^{\circ}$) & 24.09 dB & 1.90 dB & 21-29 dB \\
% 38 dBi ($2^{\circ}$) & 36.32 dB & 1.70 dB & 32-39 dB \\
% \bottomrule
% \end{tabular*}
% \label{tab:K_ant}
% \end{table}

The cumulative distribution function (CDF) of the K-factor with different antenna gains within the conference room is shown in Fig.~\ref{fig:Kf_indoor} by comparing experimental (EXP) results and curve fitting (CF). The K-factor in the dB scale is approximated by the normal distribution with mean ($\mu$) and standard deviation ($\sigma$) for different gains of antenna at the receiver. It is observable that the K-factor mean increases and variance decreases with an increase in antenna gain (i.e., with the decrease in beamwidth). Consequently, wireless THz communications and networks will require innovative adaptive link modulation and coding schemes to adapt to changing channel parameters with changing antenna gain.
%The phenomenon also indicates that with an increase in antenna gain, the scattering power to the receiver reduces while having a robust LoS link component.
Figure~\ref{fig:CF K-factor} shows the variation of the K-factor with varying antenna gains. 
% The trend shows that the K-factor increases with increasing antenna gains (i.e., decreasing antenna beam widths). 
The curve fit is given by:
\begin{equation}
    K(G) = 0.03576 G^2 - 1.246 G + 32.1,
\end{equation}
where $K$ is the Rician K-factor in dB and $G$ is the antenna gain in dBi. 
% The coefficient of determination of the fit is $R^2=0.9890$, which specifies a reasonable fit to the data points.

% \begin{figure}
%     \centering
%     \includegraphics[width=\columnwidth]{Figures/k_factor_fitting_sherif.pdf}
%     \caption{The Rician K-factor with varying antenna gains for each of the receiver locations in room 655.}
%     \label{fig:k factor fitting}
% \end{figure}

\subsubsection{Delay Spread}

To analyze the power delay profile of the indoor channel with different antenna gain, the root mean square (RMS) delay spreads (DS), $\tau_\mathrm{RMS}$, are computed for various receiver positions. The RMS DS is calculated by 
\begin{equation}\label{eq:delay_spared_in}
\tau_\mathrm{RMS} =\sqrt{\frac{\sum_{i}^{}(d_{i} - \hat{d})^2~p_{i}^{2}}{\sum_{i}^{}p_{i}^{2}}},
\end{equation}
where $d_{i}$ and $p_{i}$ are the delay and received power of propagation path $i$, respectively, and $\hat{d}$ is the mean delay represented by
\begin{equation}\label{eq:mean_delay_in}
\hat{d} =\frac{\sum_{i}^{}d_{i}~p_{i}}{\sum_{i}^{}p_{i}}.
\end{equation}

\begin{figure}
\centering
\subfloat[LoS]{\includegraphics[width =0.9\columnwidth]{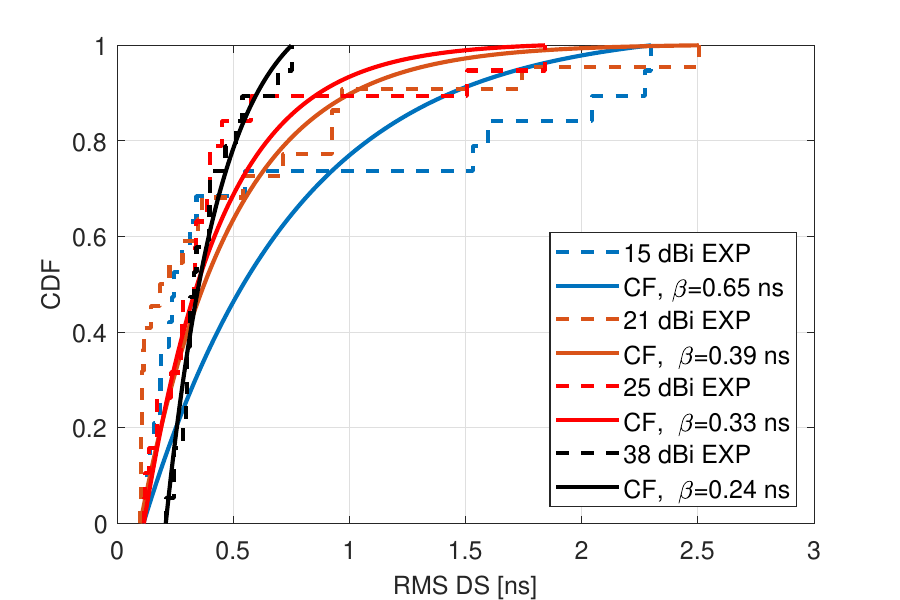}}\\
\vspace{-2mm}\subfloat[NLoS]{\includegraphics[width =0.9\columnwidth]{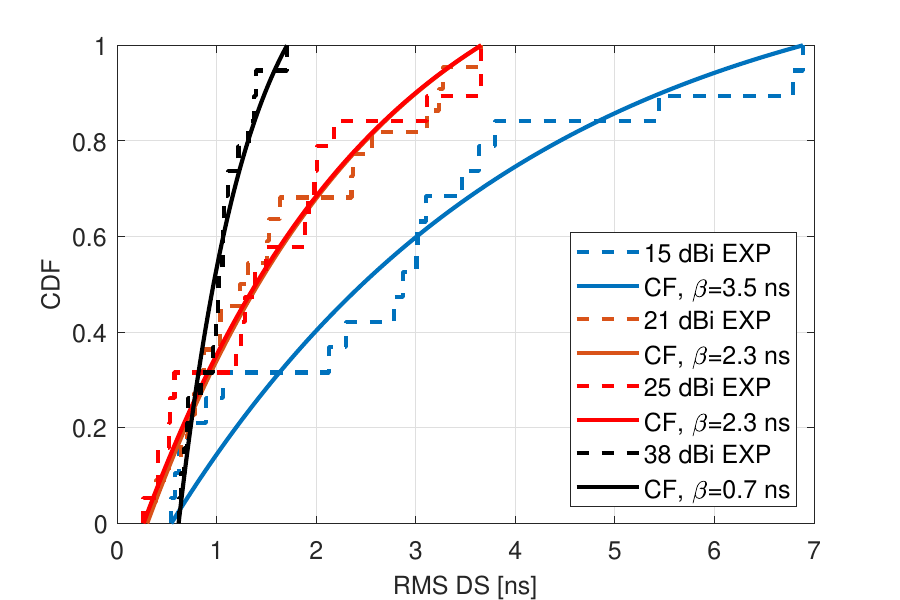}}
\caption{CDF of RMS DS with different antenna gain for EXP results (dotted line) and its CF (solid line) with $\beta$ of exponential distribution.\vspace{-1.2em}}
\label{fig:DS_indoor}
\end{figure}

% \begin{table}
% \caption{RMS DS statistics for different antenna gain in empty room 655 with transmit antenna gain 15~dBi}
% \centering
% \begin{tabular*}{\columnwidth}{@{\extracolsep{\fill}}ccc}
% \toprule
% Antenna & $\beta_\mathrm{LoS}$ [ns] & $\beta_\mathrm{NLoS}$ [ns] \\
% \midrule
% 15 dBi ($30^{\circ}$) & 0.6532 & 3.5715\\
% 21 dBi ($11^{\circ}$) & 0.3981 & 2.3008\\
% 25 dBi ($10^{\circ}$) & 0.3349 & 2.3372\\
% 38 dBi ($2^{\circ}$) & 0.2400 & 0.6964\\
% \bottomrule
% \end{tabular*}
% \label{tab:DS_ant}
% \end{table}

% \begin{figure}
% \centering
% \includegraphics[width=\columnwidth]{Figures/RMS_DS_LoS_and_NLoS_fitting_sherif.pdf}
% % \subfloat[LoS]{\includegraphics[width =0.5\textwidth]{Figures/RMS_DS_LoS_fitting_sherif.pdf}}
% % \hfill
% % \subfloat[NLoS]{\includegraphics[width =0.5\textwidth]{Figures/RMS_DS_NLoS_fitting_sherif.pdf}}
% \caption{The RMS delay spread with varying antenna gains for each of the receiver locations in room 655, both in line-of-sight and non-line-of-sight cases.}
% \label{fig:RMS delay spread}
% \end{figure}

The metric statistics is divided into two parts- LoS and NLoS, to have a holistic view considering the sparse nature of the THz channel and the significant difference between the statistics obtained from LoS and NloS components.  The CDF of the RMS DS captured for various gain antennas are shown in Fig.~\ref{fig:DS_indoor}, which can be approximated as an exponential distribution with mean value $\beta$ (increase in $\beta$ produced heavier tail CDF). 
% The details of the distributions are provided in Table~\ref{tab:DS_ant}. 
The value for the DS is notably small (in the range of 2~ns) for both the LoS and NLoS links. Although NLoS components have higher RMS DS and heavier tail distribution compared to LoS (i.e., x3 higher) counterpart. Regardless, the DS is large enough to create ISI, considering the ultra-broadband (20~GHz bandwidth) nature of the signal.
% Moreover, a clear trend can be observed in RMS DS and its $\beta$ value with changing antenna gain. Specifically, 
Further, the RMS DS increases with the increase in beam width (i.e., a decrease in antenna gain). Therefore, it is crucial to design innovative link adaptation techniques considering the statistical feature of DS based on antenna gain as well as the link condition - LoS or NLoS - when choosing pilot bits.
% This statistical feature could be utilized for efficient physical and link-layer design by developing an innovative link adaptation technique. 
For example, this aspect could be utilized to make an efficient choice for pilot bits to optimize the throughput by considering the 90 percentile point from the CDF curve of RMS DS.
% Therefore, it is necessary to develop an innovative link adaptation technique that considers DS statistics based on antenna gain as well as the link condition - LoS or NLoS - when choosing pilot bits. 
Figure~\ref{fig:CF RMS DS} shows the variation of the RMS DS with varying antenna gains, for both line-of-sight and non-line-of-sight cases. The trend shows that for both cases, the RMS delay spread decreases with increasing antenna gains (i.e., decreasing antenna beam widths), as the more directional the beam gets, the lesser multipath propagation occurs and thus the lower the delay spread. The curve fit equation is given by,
\begin{align}
    \tau_{\mathrm{RMS}_{\mathrm{LoS}}}(G) &= 0.00118 G^2 - 0.08012 G + 1.583,\\
    \tau_{\mathrm{RMS}_{\mathrm{NLoS}}}(G) &= 0.001444 G^2 - 0.1964 G + 6.101,
\end{align}
where $\tau_{\mathrm{RMS}}$ is the RMS DS in ns and $G$ is the antenna gain in dBi. 
% The coefficient of determination of fit is $R_{\mathrm{LoS}}^2=0.9907, R_{\mathrm{NLoS}}^2=0.9601$.

\subsubsection{Angular Spread}

The spatial multipath richness of the wireless channel can be determined by the angular spread (AS), which could be essential to define the link establishment possibility through NLoS. We investigate the RMS angular spread ($\mathrm{AS}_\mathrm{RMS}$) for the angle of arrival (AOA) in the azimuth direction ($\theta$) to examine the spatial multipath profile at 360-degree angles. For this purpose, we have rotated the receiver in steps of beam width in a horizontal direction. Analytically, it is represented by
\begin{equation}\label{eq:angular_spared_in}
\mathrm{AS}_\mathrm{RMS} =\sqrt{\frac{\sum_{i}^{}(\theta_{i} - \hat{\theta})^2~p_{i}^{2}}{\sum_{i}^{}p_{i}^{2}}},
\end{equation}
where $\theta_{i}$ and $p_{i}$ are the AOA and received power of propagation path $i$, respectively, and $\hat{\theta}$ is the mean AOA represented by
\begin{equation}\label{eq:mean_angle_in}
\hat{\theta} =\frac{\sum_{i}^{}\theta_{i}~p_{i}}{\sum_{i}^{}p_{i}}.
\end{equation}

\begin{figure}
\centering
\includegraphics[width=0.9\columnwidth]{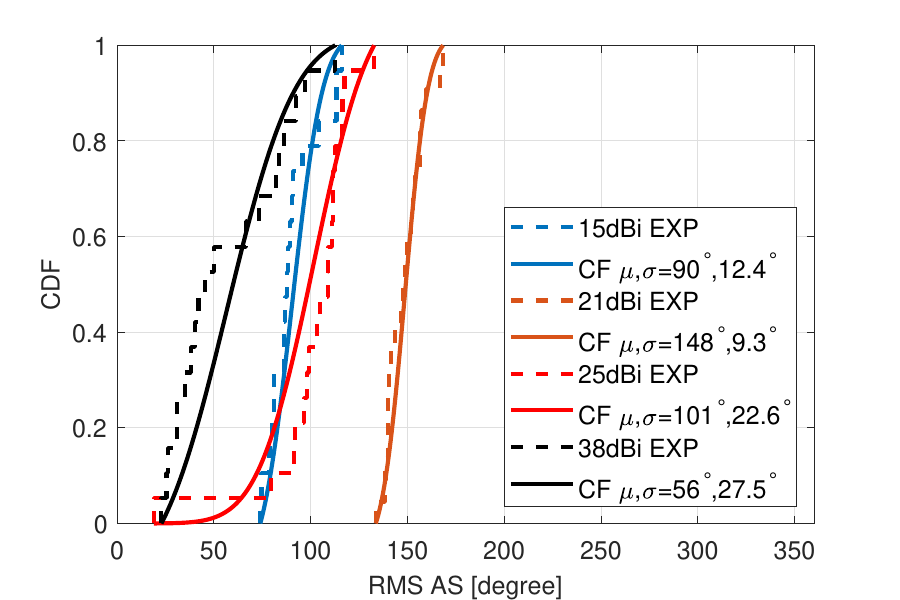}
\caption{CDF of RMS AS with different antenna gain for EXP result (dotted line) and corresponding CF (solid line) with $\mu$, and $\sigma$ of normal distribution.\vspace{-1.2em}}
\label{fig:AS_indoor}
\end{figure}

% \begin{table}
% \caption{$\mathrm{AS}_\mathrm{RMS}$ statistics for different antenna gain with transmit antenna gain of 15~dBi}
% \centering
% \begin{tabular*}{\columnwidth}{@{\extracolsep{\fill}}cccc}
% \toprule
% Antenna & $\mu$ & $\sigma$ & Total \\
% \midrule
% 15 dBi ($30^{\circ}$) & $90^{\circ}$ & $12.43^{\circ}$ & $90.33^{\circ}$ \\
% 21 dBi ($11^{\circ}$) & $148^{\circ}$ & $9.31^{\circ}$ & $154^{\circ}$ \\
% 25 dBi ($10^{\circ}$) & $101.60^{\circ}$ & $22.63^{\circ}$ & $100.60^{\circ}$ \\
% 38 dBi ($2^{\circ}$) & $56.80^{\circ}$ & $27.57^{\circ}$ & $76.70^{\circ}$ \\
% \bottomrule
% \end{tabular*}
% \label{tab:AS_ant}
% \end{table}

The CDF of the $\mathrm{AS}_\mathrm{RMS}$ with different antenna gain obtained from data captured at different positions within the room is shown in Fig.~\ref{fig:AS_indoor}. It follows the normal distribution with mean ($\mu$) and standard deviation ($\sigma$). Overall, $\mathrm{AS}_\mathrm{RMS}$  is low due to the sparse nature of NLoS paths, which indicates establishing a reliable communication link over a 360-degree angle is critical in the THz band, and it becomes increasingly challenging as antenna gain increases. It is observed by the fact that $\sigma$ increases with an increase in antenna gain.
Therefore, an extensive and innovative searching algorithm is required to establish a reliable communication link through LoS or significant NLoS components (i.e., link through a highly reflective surface). The $\mathrm{AS}_\mathrm{RMS}$ is highly dependent on the room composition and structure. Therefore, further study of the channel in different indoor scenarios of varying sizes is required to establish a correlation between $\mathrm{AS}_\mathrm{RMS}$'s statistical parameter and antenna gain.

\section{Conclusion}
\label{sec:conclusion}
This paper reports channel-sounding measurements for ultra-broadband links in the 130--150~GHz band with various antenna gains. 
% First, we describe the hardware and signal processing techniques used for the channel sounder to support measurement campaigns. Then, we analyzed the path loss model considering the near-field effect and the metrics related to the multipath profile, such as k-factor, delay spread, and angular spread. Further, the correlations between the metrics and antenna gain are shown.
The results show the impact of the antenna directivity \& beam width on the channel characteristics beyond a change in gain. Although providing environment-specific insight, these cannot be decoupled from the channel statistics through any single curve fitting or correction term. Thus, the results motivate the development of robust real-time channel estimation and link adaption techniques that consider the changes in channel properties when dynamic beams, likely to be generated from a single transmit, receive, or reflect antenna array, are utilized. This also motivated further quantization of the antenna gain in other sub-THz channel studies. Our future work is aimed at increasing the richness of the data set in additional scenarios (indoor, outdoor, aerial), with different structural and geometrical aspects, blockage, and atmospheric conditions at different frequencies above 100~GHz for a broad understanding of the sub-THz channel.

\section*{Acknowledgment}
This work is funded by the U.S. National Science Foundation under grant CNS-2225590 and the U.S. Air Force Office of Scientific Research under grant FA9550-23-1-0254. The authors would like to thank Jacob Hall, a Northeastern University graduate alumni, 
% who was at Northeastern University when the measurements were conducted, 
for his efforts in developing the control automation and data collection.
% of the NU Channel Sounder.

\bibliographystyle{IEEEtran}
\bibliography{ref}

\end{document}